# Laser optothermal nanobomb for efficient flattening of nanobubbles in van der Waals materials

Jia-Tai Huang[1], Benfeng Bai[1,*], Hong-Ren Chen[1], Peng-Yi Feng[1], Jian-Yu Zhang[1], Yu-Xiao Han[1], Xiao-Jie Wang[1], Hong-Wei Zhou[1], Yuan Chai[1], Yi Wang[1], Guan-Yao Huang[2], Hong-Bo Sun[1,*]

[1] *State Key Laboratory of Precision Measurement Technology and Instruments, Department of Precision Instrument, Tsinghua University, Beijing 100084, China*

[2] *Key Laboratory for Thermal Science and Power Engineering of Ministry of Education, Beijing Key Laboratory of CO2 Utilization and Reduction Technology, Department of Energy and Power Engineering, Tsinghua University, Beijing 100084, China.*

Corresponding to: [baibenfeng@tsinghua.edu.cn](mailto:baibenfeng@tsinghua.edu.cn); [hbsun@tsinghua.edu.cn](mailto:hbsun@tsinghua.edu.cn).

## Abstract

Nanobubbles are typical nanodefects commonly existing in two-dimensional (2D) van der Waals materials such as transition metal dioxides, especially after their transfer from growth substrate to target substrates. These nanobubbles, though tiny, may significantly alter the local electric, optoelectronic, thermal, or mechanical properties of 2D materials and therefore are rather detrimental to the constructed devices. However, there is no post-processing method so far that can effectively eliminate nanobubbles in 2D materials after their fabrication and transfer, which has been a major obstacle in the development of 2D material based devices. Here, we propose a principle, called laser optothermal nanobomb (LOTB), that can effectively flatten nanobubbles in

2D materials through a dynamic process of optothermally induced phase transition and stress-pulling effect in nanobubbles. Operation of LOTB on monolayer molybdenum disulfide (1L-MoS$_2$) films shows that the surface roughness can be reduced by more than 70% on a time scale of ~50 ms, without damage to the intrinsic property of 1L-MoS$_2$ as validated by micro-nano photoluminescence and Raman spectroscopy. Moreover, a dual-beam cascaded LOTB and a multi-shot LOTB strategies are proposed to increase the flattened area and processing effect, showing the potential of LOTB for fast nanodefect repairing in the mass production of van der Waals materials and devices.

**Introduction**

Two-dimensional (2D) van der Waals (vdW) materials, such as transition metal dioxides (TMDs),[1, 2] have emerged as important candidate materials for next-generation chip-scale optoelectronic devices.[3, 4, 5, 6] However, due to the restriction of growth at high temperatures (>500 °C)[7] and the different thermal expansion coefficients of 2D materials and growth substrates,[8] 2D materials have to be transferred from growth substrates to target substrates after growth, during which various nanodefects such as nanobubbles are inevitably generated.[9] Nanobubbles, though very tiny with dimensions typically from 10 nm to 1 μm,[10, 11] may significantly alter the dielectric environment[12, 13] and tensile strain[14, 15] of 2D materials and therefore are rather detrimental to the excitonic properties, band gap structures, and carrier mobility of the composed devices.[4, 6, 15] Some previous works have tried to reduce the impact of nanobubbles by developing clean transfer methods.[16] However, there is no post-processing method so far that can effectively eliminate nanobubbles in 2D materials, due to the vdW interactions between 2D materials and substrates.

In essence, nanobubbles are generated in 2D vdW materials due to the unexpected inclusion of impurities during their growth or transfer. Therefore, the key to remove nanobubbles is to drive away the inclusions. Previous research has shown that large-area annealing can help to improve the adhesion between TMD films and substrates.[17]

However, this method just drives small nanobubbles to merge as fewer but larger bubbles instead of eliminating them. Matthew *et al*. proposed to eliminate bubbles mechanically by pushing bubble inclusions laterally with an atomic force microscope (AFM) probe.[18] Since the probe needs to contact sample during treatment, it lacks efficiency and is only applicable to some substrates such as hexagonal boron nitride but not to hard substrates such as silica. Therefore, a high-efficiency, reliable, and versatile nanobubble elimination method is highly expected.

Here, we report an all-optical method named laser optothermal nanobomb (LOTB), which can effectively eliminate nanobubbles in 2D materials such as TMDs on a short time scale of ~50 ms by laser beam irradiation, without damage to the intrinsic optoelectronic properties of materials. The focused laser beam induces an optothermal effect on the nanobubble, which triggers the phase transition of the inclusions in nanobubbles into gas, sublimates the 2D material film to generate a tiny sacrifice breach, and then drives the inclusion to flow out of the bubble through the breach under pressure difference. As a demonstration, LOTB is employed to eliminate nanobubbles in monolayer molybdenum disulfide (1L-$MoS_2$). On this basis, by regulating the irradiated optical field, a dual-beam cascaded LOTB and a multi-shot LOTB strategies are proposed to expand the treatment areas and improve the processing effectiveness.

## Results

**Mechanics of nanobubble formation.** The formation of nanobubble in a 2D material film can be described by a modified pressurized membrane model.[19] As shown in Fig. 1a (i), a nanobubble can be regarded as the deflection of a pressurized membrane. Its profile can be expressed as

$$w(r) = h(1 - \frac{r^2}{a^2}), \tag{1}$$

where $h$ is the bubble center height, $a$ is the capping radius, and $w$ is the deflection at a radial distance of $r$ from the center. The bending energy of the membrane is negligible due to the thinness of the 2D materials.[11] The total free energy consists of the stretching energy $U_s$ and the adhesion energy $U_i$. Considering the membrane structure in equilibrium, the mechanical relationship can be determined by minimizing the total free energy ($U_s + U_i$). The aspect ratio of a nanobubble, i.e., $h/a$, can be calculated as (see details in Supporting Information, Note 1)

$$h/a \propto (\Delta\gamma / K)^{1/4}, \tag{2}$$

where $\Delta\gamma$ is the adhesion energy per unit area and $K$ is the in-plane elastic stiffness.

The mechanical and optoelectronic properties of nanobubbles are influenced by both the film-substrate interfacial interaction and the trapped inclusions inside. When nanobubbles arise after wet transfer, liquids such as water may be trapped inside the nanobubbles.[4] Previous research has shown that the adsorbed water is the most likely inclusions in nanobubbles and liquid-free interfaces of transferred 2D materials are possible only when the transfer is performed above 110 °C.[20]

For a liquid-filled nanobubble, the adhesion energies of the membrane-liquid interface, substrate-liquid interface, and membrane-substrate interface need to be considered, which can be written in the form below based on Young-Dupré equation[21]

$$\Delta\gamma = \Gamma - \gamma_c(\cos\theta_f + \cos\theta_s), \tag{3}$$

where $\Gamma$ is the adhesion energy between the membrane and substrate, $\gamma_c$ is the surface tension of the inclusion (whose value for water is around 0.078 J/m²), and $\theta_f$ and $\theta_s$ are the contact angles on the membrane and on the substrate, respectively.

For a gas-filled nanobubble, $\Delta\gamma$ can be simply taken as the adhesion energy of the membrane-substrate interface, i.e., $\Delta\gamma = \Gamma$. By taking into account Eq. (2) and Eq. (3), we know that the aspect ratio of a gas-filled nanobubble is larger than that of a liquid-filled one with the same membrane and substrate. According to the minimum total free energy principle, the pressure difference $\Delta p$ inside and outside a nanobubble can be derived as (see details in Supporting Information, Note 1)

$$\Delta p = 16(1+v)\frac{D}{a^3}\frac{h}{a} + \frac{4}{3}\frac{K}{a}\left(\frac{h}{a}\right)^3, \tag{4}$$

where $D = Et^3/12(1-v^2)$ is the flexural rigidity of the film, $E$ is the Young's modulus, $t$ is the thickness of film, and $v$ is the Poisson ratio.

**Principle of LOTB.** Based on the above mechanical analysis of nanobubbles, we propose the LOTB method whose physical process is schematically depicted in Fig. 1a. When a focused continuous-wave (CW) laser beam irradiates a single layer TMD (1L-TMD) film with liquid-filled nanobubbles after wet transfer, a light-induced temperature field is generated over the film, which consists of a high-temperature (high-temp) field region and a low-temperature (low-temp) field region. In the central region of the beam, the high-temp field sublimates the 1L-TMD film to generate a sacrifice breach so that the inclusion of the nanobubble can flow out of the bubble. While in the low-temp field region, the inclusion evaporates from liquid phase to gas phase and flows to the high-temp region under the inner pressure $p = p_0 + \Delta p$, where $p_0$ is the atmospheric pressure outside and $\Delta p$ is a pressure difference. In this way, the inclusion of nanobubble can escape from the film in the form of gas through the sacrifice breach.

As a result, the nanobubbles in the low-temp field can be flattened by paying the little price of generating a small sacrifice breach.

The optothermal process depicted in stage (ii) of Fig. 1a is justified by numerical simulation with finite element method (COMSOL Multiphysics), where an example material system of monolayer $MoS_2$ (1L-$MoS_2$) on a silica substrate is considered (see simulation details in Supporting Information, Note 2). As shown in Fig. 1b, under irradiation by a Gaussian laser beam with an intensity density of 106.1 mW/µm² and beam diameter of 0.3 µm (green line, normalized), the simulated temperature profile (black line) has a peak temperature around 900 K and a wider full width at half maximum (FWHM) compared to the laser spot due to the heat diffusion effect. By taking into account the sublimation temperature point of 1L-$MoS_2$ (723 K) and the evaporation temperature point of water (373 K), the high-temp field region (red) and low-temp field region (blue) can be identified, which have a radius of 0.18 µm ($r_H$) and a radius of 0.98 µm ($r_L$), respectively. Accordingly, the areas of the sacrifice region and flattened region can be estimated as 0.10 µm² ($S_s = \pi r_H^2$) and 2.90 µm² ($S_F = \pi r_H^2 - \pi r_L^2$), respectively. In addition to the phase transition, the stress-pulling effect induced by the expansion of nanobubbles under laser irradiation can help to flatten the film, which is discussed in detail by both simulation and Raman measurement in Supporting Information, Note 3.

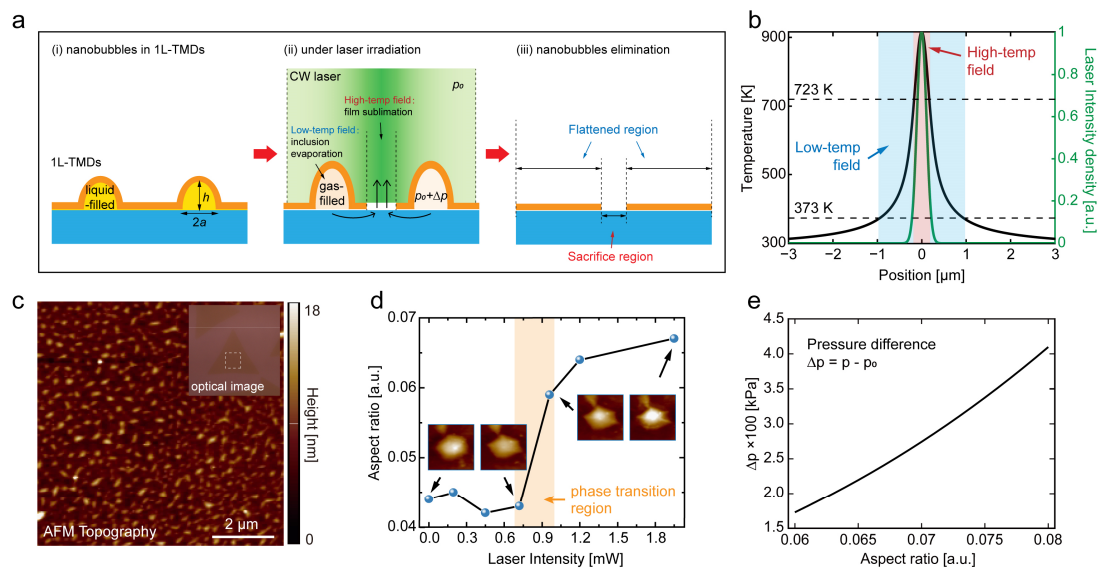

**Figure 1. Principe of LOTB.** (a) Schematic diagram of the bubble flattening process with LOTB. In stage (i), liquid-filled nanobubbles are generated during wet transfer; in stage (ii), a high-temp field region and a low-temp field region are generated under laser irradiation, which sublimates the TMD film to generate a sacrifice region and enables the phase transition of the inclusions, respectively; in stage (iii), the inclusions inside the nanobubbles flow out of the sacrifice region under the pressure difference inside and outside nanobubbles, which flattens the film together with a stress-pulling effect. (b) Simulated temperature distribution (black line) under the irradiation of a Gaussian laser beam (green line, normalized) with an intensity density of 106.1 mW/μm$^2$ and beam diameter of 0.3 μm. The red and blue regions indicate the high-temp field and low-temp field regions, respectively. (c) AFM topography of a transferred 1L-MoS$_2$ sample with nanobubbles. The inset shows the optical image of the 1L-MoS$_2$ film, where the dashed rectangle indicates the region of the AFM topography. (d) Aspect ratio change of a nanobubble under different irradiation laser power, where the orange area indicates the phase transition region leading to an abrupt change of aspect ratio. The insets show the corresponding AFM topography images of the same nanobubble under different irradiation laser powers. (e) Calculated pressure difference inside and outside the nanobubble versus the aspect ratio. The radius of the nanobubble is taken as 200 nm.

**Dynamic change of nanobubbles under laser irradiation.** To investigate the dynamic change of nanobubbles under laser irradiation experimentally, 1L-$MoS_2$ samples were prepared by chemical vapor deposition (CVD) and transferred onto glass coverslips by KOH-based wet transfer method[22] (see Methods section for details). As the AFM topography images show in Fig. 1c, many nanobubbles with radii of 50 nm ~ 400 nm and heights of 2 nm ~ 16 nm were generated throughout the film surface after transfer, which are too small to be observed by optical microscope in the inset.[4] The aspect ratio of nanobubbles can be estimated as ~0.04 according to the statistics of the nanobubble topography, similar to the case of liquid-filled blister with the material system of $MoS_2/SiO_2$ reported before.[11]

Then we illuminate the film with a focused laser beam. The wavelength of laser was selected as 532 nm, which corresponds to the side absorption band of 1L-$MoS_2$. The change of the aspect ratio of a nanobubble with a radius of around 200 nm under increasing irradiation laser power was measured by an AFM, as shown in Fig. 1d. The process is clearly divided into three stages: when the laser power is relatively low, the aspect ratio of the nanobubble changes very little; when the laser power increases to 0.7 ~ 0.96 mW, as indicated by the orange region in Fig. 1b, the aspect ratio increases sharply; and after that, the aspect ratio increases slowly. Since the laser acts as a heating source to linearly raise the temperature of the film and its inclusions with the increase of the laser power, the orange region indicates a phase transition region of the inclusions. That is, the liquid inclusion evaporates into gas so that the liquid-filled nanobubble is turned into a gas-filled one with higher aspect ratio. After that, when the laser power is further increased to 2 mW, the temperature of the gas-filled nanobubble continues to rise, leading the increase of the internal pressure and the further increase of the aspect ratio. This fact is consistent with the simulation result that the aspect ratio is nearly proportional to the temperature in a gas-filled nanobubble (see details in Supporting Information, Note 4). Moreover, according to the measured aspect ratio curve in Fig. 1d, the pressure difference $\Delta p$ inside and outside a nanobubble with respect to aspect ratio can be calculated by Eq. (4) and plotted in Fig. 1e.

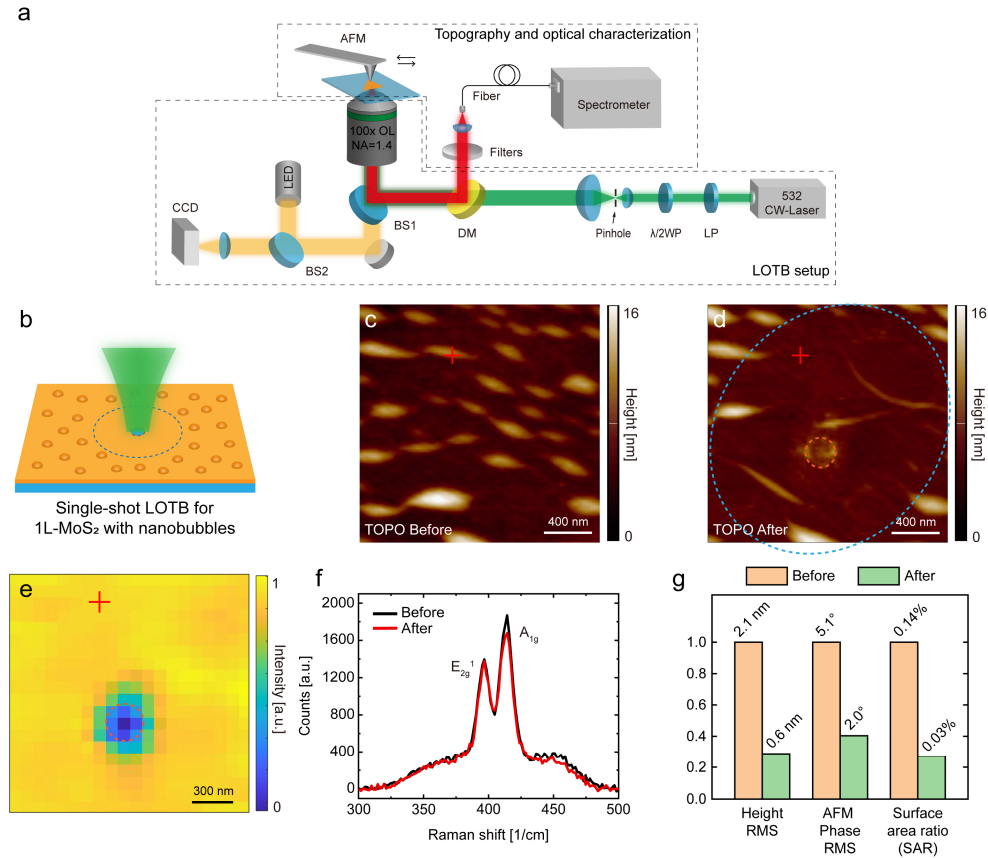

**Figure 2. Elimination of nanobubbles in 1L-MoS₂ by single-shot LOTB.** (a) Schematic of the LOTB setup, which can be used jointly with an AFM and an optical spectroscopic characterization system. (b) Schematic of nanobubble flattering on a 1L-MoS₂ sample by single-shot LOTB. (c) and (d) are AFM topography images of a 1L-MoS₂ sample with nanobubbles before and after LOTB treatment, respectively. The orange and blue dashed lines indicate the sacrifice region and the flattened region, respectively. (e) Confocal PL mapping image of the same sample area as in (d) after LOTB. (f) Raman spectra of the sample before (black curve) and after (red curve) LOTB treatment at the point marked by the red cross in (c)-(d). (g) Surface roughness statistics of the flattened region before and after LOTB.

Single-shot LOTB for nanobubble elimination. An instrumental setup has been built to implement LOTB, as schematically shown in Fig. 2a, which can be used jointly with an AFM and an optical spectroscopic characterization system. A 532 nm CW laser of 7.5 mW power is used to illuminate the sample from bottom side through an oil-immersed objective lens (100x, 1.4 NA, Olympus). The diameter of the laser spot is ~0.3 μm. The AFM probe can be aligned precisely with the laser spot. All the experiments were carried out in ambient and the detailed system information is described in Methods section.

The LOTB can be performed with single shot to eliminate the nanobubbles in 1L-MoS$_2$, as schematically shown in Fig. 2b. Figures 2c and 2d show the topography changes of the same 1L-MoS$_2$ sample area before and after LOTB treatment, respectively. Clearly, a small sacrifice region is generated in the center (orange dashed line), around which nanobubbles are eliminated and the film is obviously flattened (blue dashed line). The diameters of the sacrifice region and the flattened region are measured to be ~0.3 μm and ~2 μm, respectively, which are consistent with the simulation results in Fig. 1b. As a result, the surface roughness of the film is reduced, which can be quantified by calculating the root mean square (RMS) of topography after laser irradiation. As shown in Fig. 1g, the RMS is reduced by 71.4% (from 2.1 nm to 0.6 nm). The roughness of the sample can also be confirmed by AFM phase mapping (see details in Supporting Information, Note 5), which also decreases by 60.8% (from 5.1° to 2.0° in RMS)

Moreover, since the inclusions inside nanobubbles cause deformation of the 1L-MoS$_2$, we may define a surface area ratio $SAR = \iint (dS - S_P)/S_P$ to quantify the presence of the inclusions, where S$_P$ represents the projection area of the flattened region. The smaller the value of SAR, the closer the film is attached to the substrate without inclusions. As shown in Fig. 1g, the SAR decreases by 78.5% (from 0.14% to

0.03%) after laser irradiation, indicating that the inclusions indeed disappear instead of shifting or converging as the situation in annealing.[17] .

In order to identify whether the 1L-MoS$_2$ film was damaged in LOTB, we performed micro-nano photoluminescence (PL) and Raman spectral characterizations of the sample to investigate the change of its properties, including the optoelectric property, electronic energy band, and vibrational modes.[4, 23] The PL and Raman spectral results in Figs. 2e and 2f show that the 1L-MoS$_2$ film was intact after LOTB. Despite the PL quenching in the sacrifice region in Fig. 2e, the PL emission intensity in the flattened region is even and as strong as that in the surrounding region nearby. The Raman spectra were measured at the same point as shown in Fig. 2f, which has nearly no change in intensity and peak position, indicating that LOTB does not damage the flattened film.

The time scale of LOTB was investigated by carrying out experiments with different treatment time under the same power 6 mW through the transistor-transistor logic (TTL) modulation of the CW laser source. As shown in Fig. S4 in Supporting Information, the effect of nanobubble elimination appears at 20 ms and becomes evident after 50 ms (see details in Supporting Information, Note 6). Furthermore, to rule out the field effect and shock wave by laser, we have carried out a similar experiment using a single picosecond laser as shown in Fig. S5 in Supporting Information, which shows no flattening effect in both AFM topography and phase mapping (see details in Supporting Information, Note 7).

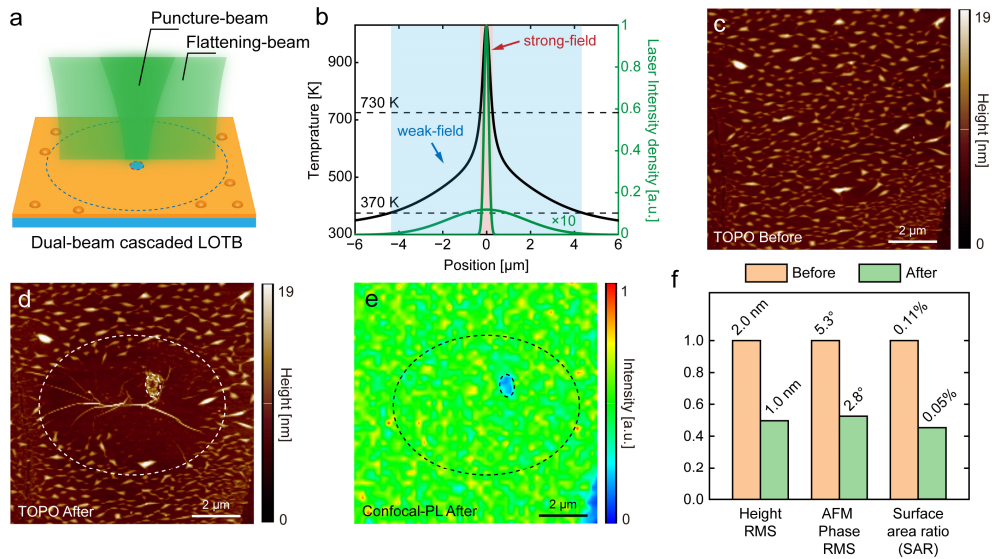

**Figure 3. Dual-beam cascaded LOTB.** (a) Schematic of implementing the dual-beam cascaded LOTB. A strong puncture-beam has a smaller diameter and high-intensity density while a relatively weak flattening-beam has a larger diameter and low-intensity density, which control the high-temp and low-temp field regions, respectively. (b) Simulated temperature distribution (black line) under the irradiation of two Gaussian lasers (green lines), consisting of a puncture-beam (with an intensity density of 106.1 mW/µm$^2$ and a spot diameter of 0.3 µm) and a flattening-beam (with an intensity density of 1.3 mW/µm$^2$ and a spot diameter of 5 µm). The red and blue regions indicate the high-temp field region and low-temp field region, respectively. (c) and (d) are AFM topography images of a 1L-MoS$_2$ sample with nanobubbles before and after the dual-beam cascaded LOTB treatment, respectively. The dashed lines indicate the sacrifice region and flattened region. (e) Confocal PL mapping images of the same sample area as that in (d) after laser irradiation. (f) Surface roughness statistics of the flattened region before and after LOTB.

**Dual-beam cascaded LOTB and multi-shot LOTB.** The single-shot LOTB uses a single Gaussian laser beam to generate both the high-temp and low-temp field regions, which limits the area and position of the flattened region. In fact, the two fields can be controlled by separate laser beams or multi-shot irradiation. In this way, we have developed a dual-beam cascaded LOTB and a multi-shot LOTB to increase the treatment area and effect.

Figure 3a shows the schematic diagram of the dual-beam cascaded LOTB, where a narrow puncture-beam and a wide flattening-beam are used to simultaneously irradiate the sample. As the simulation results in Fig. 3b show (see simulation details in Supporting Information, Note 2), the two green lines indicate the two beam profiles: one has a smaller diameter (0.3 µm) and higher intensity density (106.1 mW/µm$^2$) and one has a larger diameter (5 µm) and lower intensity density (1.3 mW/µm$^2$). While the puncture-beam generates the high-temp field region, the flattening-beam plays a role in increasing the range of the low-temp field region by decreasing the gradient of the optical and temperature fields in the center. The black line indicates the simulated in-plane temperature distribution. Compared to the distribution in Fig. 1b with single-shot LOTB, the range of the high-temp field region (red) remains nearly unchanged with a radius of 0.28 µm while the range of the low-temp field region (blue) is 4.3 times larger with a radius of 4.35 µm, determining a 20.4 times lager flattened region as 59.2 µm$^2$.

The instrumental setup of the dual-beam cascaded LOTB and the laser spot of two beams are shown in Fig. S6 in Supporting Information, Note 8. The power of the puncture-beam is 7.5 mW that is the same as the single-shot case, and the power of the flattening-beam is 25 mW. Figures 3c and 3d show the AFM topography changes of the sample film before and after LOTB treatment. The diameters of the sacrifice and flattened regions are measured to be ~0.4 um and ~6 um, consistent with the simulation results in Fig. 3b. The flattened region is around 9.2 times larger than that in the single-shot LOTB in Fig. 2. The surface roughness of the film is also investigated, as shown in Fig. 3f, where the RMS values in AFM topography and phase are decreased by 50% and 47%, respectively, and the SAR is decreased by 55%. The smaller RMS reduction can be attributed to the stress-pulling effect since the stress decays with the square of the distance in space as derived in Supporting Information, Note 3. The results also indicate that the phase transition effect is dominant in LOTB with the assistance of the stress-pulling effect. As a comparison, the nanobubbles remain unchanged when irradiating the film with only the puncture-beam as shown in Fig. S7 in Supporting Information, proving that the flattening effect in Fig. 3d is indeed the combined effect of the two beams.

To investigate the property of the treated film, confocal PL mapping was conducted and shown in Fig. 3e. Similar to the results in Fig. 2d, the confocal PL mapping reveals that the emission intensity in the flattened region is even and as strong as the surrounding area nearby, indicating that the dual-beam cascaded LOTB does not damage the flattened region.

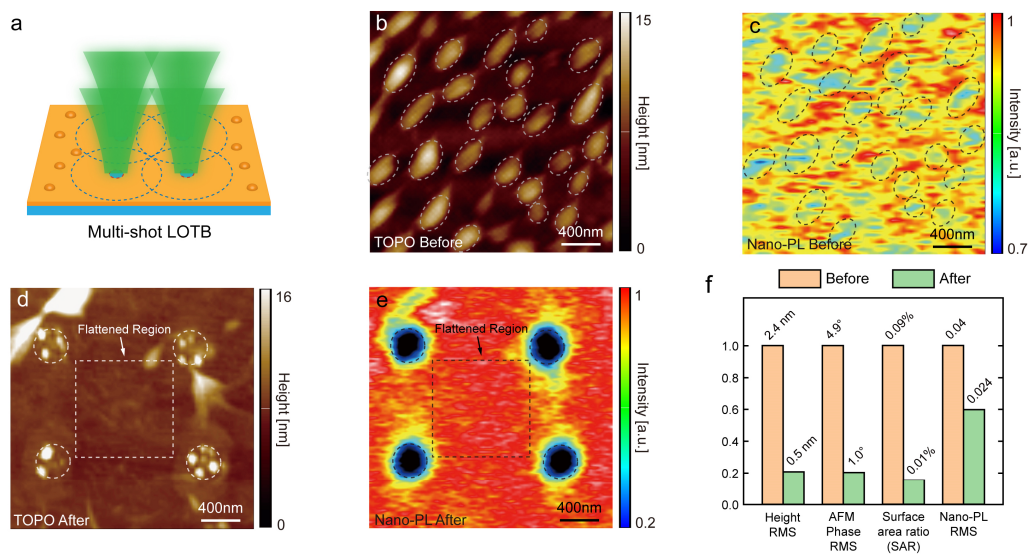

**Figure 4. Multi-shot LOTB.** (a) Schematic of the multi-shot LOTB with four lasers irradiating adjacent positions. (b) Topography and (c) nano-PL mapping images of a 1L-MoS$_2$ sample with nanobubbles. (d) AFM Topography and (e) nano-PL mapping images of the sample after four-shot LOTB with intensity density of 101.9 mW/μm$^2$ and spot diameter of 0.3 μm. The dashed cycles spaced 1.5 μm apart indicate the irradiation points and sacrifice regions under laser irradiation, while the dashed rectangle indicates the flattened region of 1 μm×1 μm. (f) Surface roughness statistics of the flattened region before and after LOTB treatment.

In addition to the dual-beam cascaded LOTB, we can also further improve the effectiveness and control the position of the treatment area by multip-shot LOTB. As a demonstration in Fig. 4a, four laser beams are used to irradiate the sample in a square array. Since each laser beam can affect a circular region centered around it, the overlapping regions in between can be controlled to realize a flattened region. As shown by the AFM topography images before and after multi-shot LOTB in Figs. 4b and 4d, the four irradiation spots (dashed circles in Figs. 4d-e) are spaced by 1.5 μm apart with power equal to 7.2 mW. The dashed rectangle indicates the flattened region of 1 μm×1 μm. According to the surface roughness statistics results in Fig. 4f, the RMS in AFM topography and phase decrease by 79.2% and 79.6%, respectively, and the SAR decreases by 88.9% after multi-shot LOTB, indicating that the flattening effect is significantly improved than that in the single-shot LOTB in Fig. 2g.

Experimental study of the optoelectronic properties of the treated region before and after LOTB is very important, but is difficult to implement at nanoscale due to the lack of experimental tool.[24] Fortunately, by applying a nano-PL microscopy that we have developed recently,[4] super-resolution mapping of the excitonic property of the TMD film can be performed in ambient condition with an ultra-high spatial resolution of 10 nm (see details of nano-PL in Methods section). As shown in Fig. 4c, before laser irradiation, the PL emission is weakened in nanobubble areas (dashed black circles in Fig. 4c), which is caused by the dielectric environment induced doping effect in nanobubbles.[25] In contrast, after multi-shot LOTB, the optoelectronic property in the flattened region (dashed black rectangle in Fig. 4e) is even and intact due to the elimination of nanobubbles, as verified by the RMS reduction in Fig. 4f. The above results show the effectiveness of LOTB to eliminate nanobubbles and improve the quanlity of 2D material based devices.

## Conclusion

In this work, we have reported an all-optical LOTB method that can effectively and efficiently flatten 1L-TMDs with nanobubbles at a time scale of ~50 ms. By theoretical simulation and experimental investigation of the sample topography during LOTB, the optothermal dynamics dominated by phase transition is revealed. By applying LOTB, nanobubbles in transferred 1L-MoS$_2$ samples were effectively eliminated with the surface roughness decreased by 71.4% and 60.8% in RMS of AFM topography and phase, respectively, and 78.5% in SAR. In addition to single-shot LOTB, we have developed dual-beam cascaded LOTB and multi-shot LOTB to expand the flattened region and improve the elimination efficiency. Moreover, the optoelectronic properties of the treated sample have been studied by confocal PL microscopy, nano-PL microscopy, and Raman spectroscopy, which show that the treated regions are intact. In the future, to meet the fabrication demand of 2D material based semiconductor devices, the LOTB treated nanobubble-free area can be as large as possible by aligning the sacrifice regions with the lines of wafer dicing.[26] Our study not only reveals the formation dynamics of nanobubbles and explores how they change under laser irradiation, but also provides an efficient and promising post-treatment tool for nanodefect repair, which may play important role in next-generation semiconductor industry.

## Methods

**LOTB system.** The LOTB system is established based on an inverted microscope (IX81, Olympus). The puncture-beam comes from a CW 532 nm laser (MGL-III-532nm, CNI) with a maximum power of 150 mW passing through a dichroic mirror with a 535 nm cut-off (FF535-SDi01, Semrock) and is focused on the sample by a 1.4 NA oil-immersion objective lens (100x, Olympus) after spatial filtering through a 50 μm pinhole (P50K, Thorlabs). The flattening-beam comes from another laser (MGL-U-532nm, CNI) with a higher power of 1.5 W. A convex lens (f = 100 mm) is used to turn the laser beam into divergence to produce a large-scale spot at the focal plane of the microscope. Then, the two laser beams are combined in one optical path through a 10:90 beam splitter. An electronic shutter (GCI-73M, Daheng Optics) is used to control the laser on and off with open time (measured time from 50%) close to 10 ms.

**AFM and nano-PL characterization.** AFM and nano-PL characterizations are performed based on a home-built near-field microscope setup combining an AFM (NTEGRA, NT-MDT) and an inverted microscope. Both the AFM probe and the sample stage are mounted on high-precision three-dimensional piezoelectric scanners (NX100, NT-MDT), with which the focused beam can be aligned with the probe and perform the sample mapping. For nano-PL characterization, the near-field PL signal under the apex is modulated by an AFM probe working in tapping mode. Then the PL signal detected by the PMT is sent to a lock-in amplifier (HF2LI, Zurich Instruments) and demodulated to extract super-resolution signal in the near field. The control of the whole system is operated with an AFM controller. The probes used in the system are commercial silicon AFM probes (NSG01, ScanSens), and the mapping results are all carried out by raster scans from left to right.

**Spectral measurement.** The PL and Raman spectral measurements are conducted by using the same setup of the LOTB system. Under the excitation laser of 532 nm wavelength, the PL and Raman signals of the sample can be collected by the same objective and filtered with a filter set including a 550 nm long-pass filter (FELH0550,

Thorlabs) and a 700 nm short-pass filter (FESH0700, Thorlabs), and then acquired by a confocal Raman spectrometer (INVA, Renishaw) or a PMT (PMT1002, Thorlabs).

**Preparation of the 1L-MoS$_2$ sample by wet transfer.** Commercial 1L-MoS$_2$ films are grown by CVD on Al$_2$O$_3$ substrates and then transferred to the desired substrates (here, glass coverslips) by KOH-based wet transfer method. Specifically, the 1L-WS$_2$ sample on the initial substrate with polymethyl methacrylate (PMMA) is heated at 100 °C, and then immersed in 2 mol/L KOH solution for several hours until the PMMA film is separated from the substrate naturally. After that, the PMMA film is washed by deionized water for 3-5 times and transferred onto glass coverslip. Then the sample is left to dry naturally and heated at 80°C. Finally, to etch the PMMA, the sample is immersed in acetone for 3-5 times with 30 minutes for each time. The growth and transfer are both performed by 6Carbon Technology (Shenzhen).

## Corresponding Author

**Benfeng Bai**. *State Key Laboratory of Precision Measurement Technology and Instruments, Department of Precision Instrument, Tsinghua University, Beijing 100084, China; https://orcid.org/0000-0002-5348-9589;*

Email: baibenfeng@tsinghua.edu.cn

**Hong-Bo Sun**. *State Key Laboratory of Precision Measurement Technology and Instruments, Department of Precision Instrument, Tsinghua University, Beijing 100084, China; https://orcid.org/0000-0003-2127-8610;*

Email: hbsun@tsinghua.edu.cn

## Notes

The authors declare no competing financial interest.

# ACKNOWLEDGMENT

The authors acknowledge the financial support of the National Natural Science Foundation of China (62175121, 61960206003).